\documentclass[preprint, amsmath,amssymb,aps]{revtex4-1}
\usepackage{graphicx}
\usepackage{dcolumn}
\usepackage{bm}
\usepackage[pdftex,colorlinks]{hyperref}
\hypersetup{citecolor=blue}
\begin{document}

\preprint{APS/123-QED}

\title{Quantization of surface plasmon polariton by Green's tensor method in amplifying and attenuating media }
\author{Z. Allameh}
\affiliation{Institute of Laser and Plasma, Shahid Beheshti University, Tehran, Iran\\}
 \author{R. Roknizadeh}
 \affiliation{
 Department of Physics, Quantum Optics Group\\University of Isfahan, Isfahan, Iran}
\author{R. Masoudi}
\affiliation{
 Institute of Laser and Plasma, Shahid Beheshti University, Tehran, Iran\\}
 
\date{\today}

\begin{abstract}
In this paper we will present a quantization method for  SPP (Surface Plasmon Polariton) based on Green's tensor method, which is applied usually for  quantization of EM field  in various dielectric media. This method will be applied for a semi-infinite structure, which contains metal and dielectric regions with one interface.
Moreover, by introducing the  quantized SPP, we will investigate the SPP propagation
in the attenuating and amplifying systems. We will also consider two modes 
of SPP, \emph{i.e.},  coherent and squeezed states, and  finally compare the propagation of these modes in the amplifying media.
\begin{description}
\item[PACS]  42.50.-p 
\end{description}
\end{abstract}

\pacs{Valid PACS appear here}
\maketitle

\section{introduction}
A SPP is a an electromagnetic 2-dimensional excitation existing on the metal-dielectric interface, which decays exponentially with diffusing depth in the metal. The term ''surface plasmon''  refers to the collective electrons oscillation on the metal surface ,  and 
''polariton'', which combine exciton and photon properties as  a quasi-particle, indicates  that the SPP involves coupling of electromagnetic waves and  the dielectric excitations.  Due to the unique properties of the SPP,  such as capacity of confining the  electromagnetic field \cite{1,2}, it  has become one of the interesting field of applied research \cite{3}.
 
By discovering the  quantum behavior of SPP in  theoretical and experimental investigations \cite{4,5,6,7}, the possibility of using them in  quantum technology,  such as Quantum Information Processing (QIP) \cite{8,9,10} and  generating  and controlling plasmonic excitation at the quantum level,  has been provided \cite{11,12}.

In order to describe the plasmonic quantum mechanically, a quantization method of SPP  is introduced  in \cite{13}. This formalism is based on Hopfield approach and  therefore  do not consider the dissipation within  the  medium. 
It is well known that   the real  part of permittivity  is related, by the Kramers-Kronig relation, to the stored energy or dispersion  within medium, and imaginary part yields the dissipation (or loss) of the energy within it.  For lossy and dispersive dielectric, Huttner  and  Barnett  have proposed a  microscopic quantization method \cite{14} . In  their method the absorptive medium  is assumed to be connected to a reservoir and represented by a collection of interacting matter fields. \\
As another most interesting scheme for quantization, the  Green’s function method   have been  proposed and  used by many authors \cite{15,16,17,18}. The Green’s function is constructed for  classical Maxwell equations  and contains, as a whole,   the system and the source of the  noise  associated with the absorption of radiation in it \cite{19,20,21,22}.

  In this contribution  we will try to apply this method for quantization of  SPP field. 
Moreover, we apply this approach to  propagation of the SPP in the amplifying media. Nowadays, because of the significant applications of this field, many efforts have been devoted to the SPP amplification \cite{32,33,34,35,36,37}.The quantum approach   not only enables  us  to take into account the SPP amplification,  but also the behavior of SPP's different states can be distinguished \cite{38}. 

The paper is organized as follows. In section 2 we review the quantization of EM-field  in dielectric media.  Regarding to  the SPP propagation along the interface of metal-dielectric, we study  a  single flat interface  system in section 3. In section 4 we apply this approach for investigating  the SPP propagation in attenuating and amplifying system  for  coherent and squeezed states  of SPP.
 A  conclusions  is given in sec 5.
 \section{Basic concepts}
 In order to obtain a suitable form of the Maxwell equations for quantization procedure, we review the basic concepts in classical and quantum electrodynamics \cite{15,16,17,18}.\\
 At first, it is useful to express electromagnetic fields in frequency domain, and separate them in negative and positive frequency components. For instance the electric field is written in the form, 
 \begin{equation}
 E(r,\omega)= E^{+}(r,\omega)+ E^{-}(r,\omega),\label{1}
 \end{equation}
 here
\begin{equation}
E^{+}(r,t)=\dfrac{1}{\sqrt{2\pi}}\int_{0}^{+\infty}\text{d}\omega E^{+}(r,\omega)\exp(-i\omega t)\label{2}
\end{equation} 
and the other  vector fields are expressed analogously.  Then, in a  medium without free charges and currents, the classical maxwell equations for positive part in the frequency domain are given by, 
\begin{align}
\nabla\cdot D^{+}(r,\omega)&=0,\nonumber\\
\nabla\cdot B^{+}(r,\omega)&=0,\nonumber\\
\nabla\times E^{+}(r,\omega)&=i\omega B^{+}(r,\omega) ,\nonumber\\
\nabla\times H^{+}(r,\omega)&=- i\omega D^{+}(r,\omega) ,\label{3}
\end{align}
where $ D^{+}(r,\omega) $ and $ B^{+}(r,\omega) $ satisfy the constitutive relations:
\begin{align}
D^{+}(r,\omega)&=\epsilon_{0}\epsilon(r,\omega)E^{+}(r,\omega),\label{4}\\
B^{+}(r,\omega)&=\mu_{0} H^{+}(r,\omega).\label{5}
\end{align}
where $ \epsilon(r,\omega) $ is complex dielectric function, its real and imaginary parts  are related  by Kramers- Kronig relation. Also, $ \epsilon(r,\omega) $ is related to complex refractive index, $ n(r,\omega)$,
\begin{equation}
\epsilon(r,\omega)=[n(r,\omega)]^{2},\label{6}
\end{equation}
where
\begin{equation}
n(r,\omega)=\eta(r,\omega)+i\kappa(r,\omega).\label{7}
\end{equation}
In frequency domain, we have the following relations,  
\begin{equation}
\epsilon(r,-\omega)=\epsilon(r,\omega)^{*}, \qquad n(r,-\omega)=n(r,\omega)^{*}.\label{8}
\end{equation}
As a general property for $ \epsilon(r,\omega) $ it is, in the whole upper half of the complex frequency plane,  an analytic function without zeros. So that, with respect to the  causality, $ \epsilon(r,\omega) $ has asymptotic behavior for high frequency \cite{23,24}, 
\begin{equation}
\epsilon(r,\omega)\rightarrow 1, \quad\text{for} \quad\omega\rightarrow \infty \label{9}
\end{equation} 
In quantizing the EM-vector fields and constitutive relations \eqref{4} and \eqref{5}, we have to take into account the fluctuation- dissipation theorem. According to this theorem, an unavoidable consequence of the radiation absorption (or dissipation)  in reservoir,  is injection the additional noise (like thermal radiation) from environment to the fields.  
The noise term can be expressed  by introducing a polarization operator in  Eq. \eqref{4}, 
\begin{equation}
\hat{D}^{+}(r,\omega)=\epsilon_{0}\epsilon(r,\omega)\hat{E}^{+}(r,\omega)+\hat{P}^{+}_{N}(r,\omega).\label{10}
\end{equation}
So the quantized Maxwell equations are obtained as:\\
\begin{align}
&\nabla\cdot [\epsilon_{0}\epsilon(r,\omega) \hat{E}^{+}(r,\omega)]=\hat{\rho}^{+}_{N}(r,\omega),\label{11}\\
&\nabla\cdot \hat{B}^{+}(r,\omega)=0,\label{12}\\
&\nabla\times \hat{E}^{+}(r,\omega)=i\omega \hat{B}^{+}(r,\omega) ,\label{13}\\
&\nabla\times \hat{B}^{+}(r,\omega)=- i\omega \mu_{0}\epsilon_{0}\epsilon(r,\omega) \hat{E}^{+}(r,\omega)+\mu_{0}\hat{j}^{+}_{N}(r,\omega),\label{14}
\end{align}
here $ \hat{\rho}^{+}_{N}(r,\omega) $ and $ \hat{j}^{+}_{N}(r,\omega) $ are noise charge density and noise current density respectively which are related to $ \hat{P}^{+}_{N}(r,\omega) $ by, 
\begin{align}
\hat{\rho}^{+}_{N}(r,\omega) &=- \nabla\cdot\hat{P}^{+}_{N}(r,\omega), \nonumber\\
\hat{j}^{+}_{N}(r,\omega) &=- i\omega\hat{P}^{+}_{N}(r,\omega).\label{15}
\end{align}
It is suitable to obtain a partial differential equation for $ \hat{A}(r,\omega) $.  By expressing the field operators in terms of vector potential operator $ \hat{A}(r,\omega) $:
\begin{align}
\hat{E}^{+}(r,\omega)&=i\omega\hat{A}^{+}(r,\omega),\nonumber\\
\hat{B}^{+}(r,\omega)&=\nabla\times\hat{A}^{+}(r,\omega).\label{16}
\end{align}
and applying Eqs. \eqref{13} and \eqref{14}, one obtains:
\begin{equation}
-\nabla\times\nabla\times \hat{A}^{+}(r,\omega)+\dfrac{\omega^{2}}{c^{2}}\epsilon(r,\omega)\hat{A}^{+}(r,\omega)=-\mu_{0}\hat{j}_{N}^{+}(r,\omega).\label{17}
\end{equation}
A standard method  for solving  this equation is using the  Green's tensor, 
\begin{equation}
\hat{A}^{+}(r,\omega)=-\mu_{0}\int_{-\infty}^{+\infty}\text{d}r^{'}G(r,r^{'},\omega).\hat{j}_{N}^{+}(r^{'},\omega),\label{18}
\end{equation}
and therefore it  has to solve the following equation, 
\begin{equation}
-\nabla\times\nabla\times G(r,r^{'},\omega)+\dfrac{\omega^{2}}{c^{2}}\epsilon(r,\omega)G(r,r^{'},\omega)=I \delta(r-r^{'}).\label{19}
\end{equation}
where $ I $ is the unit tensor. The Green's tensor has  some general properties, which are   given in appendix A. 
 In the field quantization one requires the vector fields satisfying the  canonical commutation relations at any point of space, 
\begin{equation}
[\hat{A}_{i}(r,t), -\epsilon_{0}\hat{E}_{j}(r^{'},t)]=i\hbar \delta(r-r^{'})\delta_{ij}.\label{20}
\end{equation}
The main reason for accuracy of  this relation in absorbing media is the noise sources. The commutation relation for the noise current operators is given by:
\begin{align}
[\hat{j}_{N}^{(+)}(r,\omega), \hat{j}_{N}^{(-)}(r^{'},\omega^{'})]&=\vert\alpha(\omega)\vert \delta(r-r^{'})\delta(\omega -\omega^{'}),\nonumber\\
[\hat{j}_{N}^{(+)}(r,\omega), \hat{j}_{N}^{(+)}(r^{'},\omega^{'})]&=[\hat{j}_{N}^{(-)}(r,\omega), \hat{j}_{N}^{(-)}(r^{'},\omega^{'})]=0.\label{21}
\end{align} 
where $ \alpha(\omega) $ is determined such that Eqs.\eqref{20} and \eqref{21}  be consistent with each other. On the other hand the noise current is related to Langevin operator,
\begin{equation*}
\hat{j}_{N}^{(+)}(r,\omega)=\sqrt{\alpha(\omega)}\hat{f}(r,\omega).
\end{equation*}
 Also one can obtain the propagators of noise current \cite{15,16,17}, 
\begin{align}
&\langle 0\vert\hat{j}_{N}^{(+)}(r,\omega)\hat{j}_{N}^{(+)}(r^{'},\omega^{'})\vert 0\rangle =\langle 0\vert\hat{j}_{N}^{(-)}(r,\omega)\hat{j}_{N}^{(-)}(r^{'},\omega^{'})\vert 0\rangle =0\nonumber\\
&\langle 0\vert\hat{j}_{N}^{(-)}(r,\omega)\hat{j}_{N}^{(+)}(r^{'},\omega^{'})\vert 0\rangle =0\nonumber\\
&\langle 0\vert\hat{j}^{(+)}_{N}(r,\omega)\hat{j}_{N}^{(-)}(r^{'},\omega^{'})\vert 0\rangle =\vert\alpha(\omega)\vert \delta(r-r^{'})\delta(\omega -\omega^{'})\label{21a}
\end{align}  
\section{Quantization of the SPP field on a metal- dielectric interface}
\begin{figure}[ht]
\begin{center}
\includegraphics[scale=0.5]{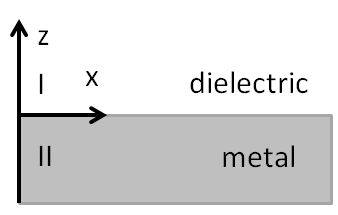}
\end{center}
\caption{Semi-infinite media, SPP propagates along  the interface  of metal- dielectric.}\label{f1}
\end{figure}

In this section we consider the quantized field of  SPP, which  propagates along the metal-dielectric interface lying 
in $xy$-plane, as shown schematically in Fig.\ref{f1}.
 For simplicity we assume that the both media  are homogeneous,  so that the related dielectric constants are independent of space points. The dielectric constant corresponding  to the whole  space is written as, 
\begin{equation}
\epsilon(r,\omega)=\epsilon_{m}(\omega)\Theta(-z)+\epsilon_{d}(\omega)\Theta(z),\label{21b}
\end{equation}
 here $ \epsilon_{m}(\omega) $ and $ \epsilon_{d}(\omega) $ are dielectric functions  of metal and dielectric media, respectively and  $ \Theta $ is the step function. 
In order to use the Eqs.\eqref{17} and \eqref{18} as starting points  for quantization process, we must derive  the Green's tensor.
\subsection{Construction of the Green's tensor}
For construction of  the Green's tensor we  use the method of  eigenmode expansion \cite{25,26,27}. In this approach, we consider the generalized form of Eq.\eqref{17}, which contains a set of eigenmodes  and eigenvalues $ (A_{n}, \lambda_{n}) $,
\begin{equation}
-\nabla\times\nabla\times A_{n}(r)+\dfrac{\omega^{2}}{c^{2}}\epsilon(r,\omega)A_{n}(r)=\epsilon(r,\omega)\lambda_{n}A_{n}(r).\label{22}
\end{equation}
These eigenmodes satisfy the orthogonality condition:
\begin{equation}
\int_{-\infty}^{+\infty}\epsilon(r,\omega) A_{n}(r)\cdot [A_{m}(r)]^{*} \text{d}^{3}r=\text{N}_{n}\delta_{nm}.\label{23}
\end{equation}
By obtaining the explicit form of eigenmodes and calculating the eigenvalues and normalization factor from Eqs. \eqref{22} and \eqref{23}, the Green's tensor is given by, 
\begin{equation}
G(r,r^{'})=\sum_{n}\dfrac{A_{n}(r) [A_{n}(r^{'})]^{*}}{\text{N}_{n}\lambda_{n}}.\label{24}
\end{equation}
By this method the Green's tensor, for geometry shown in Fig \ref{f1} can be derived. The SPP's eigenmodes can be written, 
\begin{align}
A_{k_{x}}(x,z)&=(i\hat{x}+\dfrac{ k_{x}}{\nu_{i}}\hat{z})e^{(\nu_{i}z)}e^{ik_{x}x},\quad z<0\nonumber\\
A_{k_{x}}(x,z)&=(i\hat{x}-\dfrac{ k_{x}}{\nu_{0}}\hat{z})e^{(-\nu_{0}z)}e^{ik_{x}x},\quad z>0\label{25}
\end{align} 
here $ \nu_{i} $ and $ \nu_{0} $ are out of plane decay constant and are given by:
\begin{align}
\nu_{i}&=\sqrt{\dfrac{-\epsilon_{m}}{\epsilon_{d}}}\vert k_{x}\vert ,\nonumber\\
\nu_{0}&=\sqrt{\dfrac{-\epsilon_{d}}{\epsilon_{m}}}\vert k_{x}\vert .\label{26}
\end{align}
$ k_{x} $ is the in plane propagation wave vector. The details of this calculation have been given in appendix A.
By inserting Eq.\eqref{25} in to Eqs. \eqref{22} and \eqref{23}, $ \lambda_{n} $ and $ N_{n} $ are obtained:
\begin{align}
\lambda_{n}&=k_{0}^{2}-k_{x}^{2}\dfrac{\epsilon_{d}+\epsilon_{m}}{\epsilon_{m}\epsilon_{d}},\nonumber\\
N_{n}&=\dfrac{\pi}{\vert k_{x}\vert}\sqrt{\dfrac{-\epsilon_{m}}{\epsilon_{d}}}\dfrac{(\epsilon_{d}^{2}-\epsilon_{m}^{2})}{\epsilon_{m}}.\label{27}
\end{align}
According to Eqs. \eqref{24} and \eqref{25}, for $z,z^{'}<0$ and $ z>0, z^{'}<0 $,  the Green's tensor can be written as:
\begin{align}
&G(r,r',\omega_{0})=- C\int_{-\infty}^{\infty}\text{d}k_{x}\lbrace\dfrac{\vert k_{x}\vert}{(k_{spp}^{2}-k_{x}^{2}-i0^{+})}\nonumber\\
&[(i\hat{x}+\dfrac{ k_{x}}{\nu_{i}}\hat{z})e^{(\nu_{i}z)}\Theta(-z)+(i\hat{x}-\dfrac{ k_{x}}{\nu_{0}}\hat{z})e^{(-\nu_{0}z)}\Theta(z)]\times\nonumber\\
&[(i\hat{x}+\dfrac{ k_{x}}{\nu_{i}}\hat{z})e^{(\nu_{i}z^{'})}\Theta(-z^{'})+(i\hat{x}-\dfrac{ k_{x}}{\nu_{0}}\hat{z})e^{(-\nu_{0}z^{'})}\Theta(z^{'})]\nonumber\\
&e^{ik_{x}(x-x^{'})}\rbrace , \label{28}
\end{align}
 here $ \Theta $ is the step function and $ i0^{+} $ in the denominator is necessary for obtaining the retarded Green's tensor and:
\begin{align}
C&=\dfrac{\epsilon_{m}\epsilon_{d}}{(\epsilon_{d}+\epsilon_{m})\pi\sqrt{\dfrac{-\epsilon_{m}}{\epsilon_{d}}}\dfrac{(\epsilon_{d}^{2}-\epsilon_{m}^{2})}{\epsilon_{m}}},\label{29}\\
k^{2}_{spp}&=k^{2}_{0}\dfrac{\epsilon_{m}\epsilon_{d}}{\epsilon_{d}+\epsilon_{m}}=(k^{'}+k^{"})^{2}.\label{30}
\end{align} 
By using  that $ \text{lim}_{\epsilon\rightarrow 0}\dfrac{1}{y -i\epsilon}=PV\dfrac{1}{y}+i\pi\delta(y)$ and the residu theorem one can evaluate the integral in Eq.\eqref{28}. This result in the  far- field approximation yields:
\begin{align}
&G(r,r',\omega_{0})=-i C^{'}\times\nonumber\\
&[(i\hat{x}+\dfrac{ k_{spp}}{\nu_{i}}\hat{z})e^{(\nu_{i}z)}\Theta(-z)+(i\hat{x}-\dfrac{ k_{spp}}{\nu_{0}}\hat{z})e^{(-\nu_{0}z)}\Theta(z)]\times\nonumber\\
&[(i\hat{x}+\dfrac{ k_{spp}}{\nu_{i}}\hat{z})e^{(\nu_{i}z^{'})}\Theta(-z^{'})+(i\hat{x}-\dfrac{ k_{spp}}{\nu_{0}}\hat{z})e^{(-\nu_{0}z^{'})}\Theta(z^{'})]\nonumber\\
&e^{ ik_{spp}\vert x-x^{'}\vert} , \label{31}
\end{align}
where $ C^{'}=\pi C $.
\subsection{Field quantization and Canonical commutation relations}
By substitution the  Green's tensor Eq. \eqref{31} into the  Eq. \eqref{18}, the vector potential operator is now obtained, 
\begin{align}
&\hat{A}^{+}(r,\omega)=i\mu_{0}C^{'}\times\nonumber\\
&[(i\hat{x}+\dfrac{ k_{spp}}{\nu_{i}}\hat{z})e^{(\nu_{i}z)}\Theta(-z)+(i\hat{x}-\dfrac{ k_{spp}}{\nu_{0}}\hat{z})e^{(-\nu_{0}z)}\Theta(z)] \times\nonumber\\
&\int_{-\infty}^{\infty}\int_{-\infty}^{\infty}[\text{d}z^{'}\text{d}x^{'}e^{ik_{spp}\vert x-x^{'}\vert}\hat{j}_{N}^{+}(x^{'},z^{'},\omega)\cdot\nonumber\\
&[(i\hat{x}+\dfrac{ k_{spp}}{\nu_{i}}\hat{z})e^{(\nu_{i}z^{'})}\Theta(-z^{'})+(i\hat{x}-\dfrac{ k_{spp}}{\nu_{0}}\hat{z})e^{(-\nu_{0}z^{'})}\Theta(z^{'})]. \label{32}
\end{align}
The noise current contains two components, which are related to two regions (metal and dielectric), 
\begin{align}
\hat{j}_{N}^{(+)}(x,z,\omega)&=\hat{j}_{N}^{(+)m}(x,z,\omega)\Theta(-z)+\hat{j}_{N}^{(+)d}(x,z,\omega)\Theta(z)\nonumber\\
&=(\sqrt{\alpha^{m}(\omega)}\Theta(-z)+\sqrt{\alpha^{d}(\omega)}\Theta(z))\hat{f}(x,z,\omega).\label{32a}
\end{align}
It is convenient to separate the vector potential operator into two components, rightwards and leftwards, based on the propagation direction,
\begin{align}
&\hat{A}^{+}(r,\omega)=i(\dfrac{ \beta(\omega)}{2k^{"}})^{\frac{1}{2}}\mu_{0}C^{'}\times\nonumber\\
&\lbrace(i\hat{x}+\dfrac{ k_{spp}}{\nu_{i}}\hat{z})e^{(\nu_{i}z)}\Theta(-z)+(i\hat{x}-\dfrac{ k_{spp}                  }{\nu_{0}}\hat{z})e^{(-\nu_{0}z)}\Theta(z)\rbrace \times\nonumber\\
&[\hat{a}_{R}(x,\omega)+\hat{a}_{L}(x,\omega)], \label{33}
\end{align}
where $ \hat{a}_{R} ( \hat{a}_{L}) $ is annihilation operator for rightwards (leftwards)  SPP modes and have the explicit form, 
\begin{align}
& \hat{a}_{R}(x,\omega)=(\dfrac{2k^{"}}{\beta(\omega)})^{\frac{1}{2}} \int_{-\infty}^{\infty}\int_{-\infty}^{x}\text{d}x^{'}\text{d}z^{'} e^{ik_{spp}( x-x^{'})}\times\nonumber\\
&[(i\hat{x}+\dfrac{ k_{spp}}{\nu_{i}}\hat{z})e^{(\nu_{i}z^{'})}\cdot\hat{j}_{N}^{(+)m}(x,z,\omega)\Theta(-z)\Theta(-z^{'})+\nonumber\\
&(i\hat{x}-\dfrac{ k_{spp}}{\nu_{0}}\hat{z})e^{(-\nu_{0}z^{'})}\cdot\hat{j}_{N}^{(+)d}(x,z,\omega)\Theta(z)\Theta(z^{'})], \nonumber\\
& \hat{a}_{L}(x,\omega)=(\dfrac{2k^{"}}{\beta(\omega)})^{\frac{1}{2}} \int_{-\infty}^{\infty}\int_{x}^{\infty}\text{d}x^{'}\text{d}z^{'} e^{-ik_{spp}( x-x^{'})} \times\nonumber\\
&[(i\hat{x}+\dfrac{ k_{spp}}{\nu_{i}}\hat{z})e^{(\nu_{i}z^{'})}\cdot\hat{j}_{N}^{(+)m}(x,z,\omega)\Theta(-z)\Theta(-z^{'})+\nonumber\\
&(i\hat{x}-\dfrac{ k_{spp}}{\nu_{0}}\hat{z})e^{(-\nu_{0}z^{'})}\cdot\hat{j}_{N}^{(+)d}(x,z,\omega)\Theta(z)\Theta(z^{'})],\label{34}
\end{align}
where
\begin{equation*}
\beta(\omega)=(1+\dfrac{\vert k_{spp}\vert^{2}}{\vert\nu_{i}\vert^{2}})\dfrac{\vert\alpha^{m}(\omega)\vert}{\nu_{i}+\nu_{i}^{*}}+(1+\dfrac{\vert k_{spp}\vert^{2}}{\vert\nu_{0}\vert^{2}})\dfrac{\vert\alpha^{d}(\omega)\vert}{\nu_{0}+\nu_{0}^{*}}
\end{equation*}
By this definition, one can obtain the bosonic commutation relation at the same positions, 
\begin{align}
&[\hat{a}_{R}(x,\omega),\hat{a}^{\dagger}_{R}(x,\omega^{'})]=[\hat{a}_{L}(x,\omega),\hat{a}^{\dagger}_{L}(x,\omega^{'})]=\delta(\omega -\omega^{'}),\nonumber\\
&[\hat{a}_{R}(x,\omega),\hat{a}^{\dagger}_{L}(x,\omega^{'})]=[\hat{a}_{L}(x,\omega),\hat{a}^{\dagger}_{R}(x,\omega^{'})]=0.\label{35}
\end{align}
The representation of the vector potential in quantized scheme Eq.\eqref{33} is acceptable when the canonical commutation relation \eqref{20} is satisfied simultaneously. In order to show  the accuracy of the relation \eqref{20}, it is advantageous  to derive the commutation relations for the creation and annihilation operators at any two points of space along the $ x $ direction, 
\begin{align}
&[\hat{a}_{R}(x,\omega),\hat{a}^{\dagger}_{R}(x^{'},\omega^{'})]=[\hat{a}_{L}(x^{'},\omega^{'}),\hat{a}^{\dagger}_{L}(x,\omega)]=\nonumber\\
&\delta(\omega -\omega^{'})\exp (ik^{'}(x-x^{'}))\exp( -k^{"}\vert x-x^{'}\vert ),\label{36}\\
&[\hat{a}_{R}(x,\omega),\hat{a}^{\dagger}_{L}(x^{'},\omega^{'})]=[\hat{a}_{L}(x^{'},\omega^{'}),\hat{a}^{\dagger}_{R}(x,\omega)]=\nonumber\\
&\delta(\omega - \omega^{'})\Theta(x-x^{'})\dfrac{2k^{"}}{k^{'}}\exp(-k^{"}(x-x^{'}))\sin k^{'}(x-x^{'}).\label{37}
\end{align}
The relations \eqref{36} and \eqref{37} are the same such as evaluated in Ref. \cite{15} for field operator in dielectric medium. Matloob  \emph{et al.} in \cite{15}, have interpreted these relations, and also  mentioned that the presence of the loss leads to coupling the rightwards and leftwards operators. By applying these relations,  the canonical commutation relation is written in the following form. 
\begin{align}
&[\hat{A}(r,t), -\epsilon_{0}\hat{E}(r^{'},t)]=\int_{0}^{\infty}\text{d}\omega \vert C^{'}\vert^{2}\mu_{0}^{2}\dfrac{i\omega\epsilon_{0}\beta(\omega)}{2k^{"}\pi}\times\nonumber\\
&\dfrac{\vert k_{spp}\vert^{2}}{k^{'}}\lbrace\dfrac{e^{ik_{spp}\vert x-x^{'}\vert}}{k_{spp}}+\dfrac{e^{-ik^{*}_{spp}\vert x-x^{'}\vert}}{k^{*}_{spp}}\rbrace\times\nonumber\\
&[(i\hat{x}+\dfrac{ k_{spp}}{\nu_{i}}\hat{z})e^{(\nu_{i}z)}\Theta(-z)+(i\hat{x}-\dfrac{ k_{spp}}{\nu_{0}}\hat{z})e^{(-\nu_{0}z)}\Theta(z)] \times\nonumber\\
&[(-i\hat{x}+\dfrac{ k^{*}_{spp}}{\nu^{*}_{i}}\hat{z})e^{\nu^{*}_{i}z^{'}}\Theta(-z^{'})+(-i\hat{x}-\dfrac{ k^{*}_{spp}}{\nu^{*}_{0}}\hat{z})e^{-\nu^{*}_{0}z^{'}}\Theta(z^{'})].\label{38} 
\end{align}
By some calculations, done in appendix B, the convenient form of the above equation can be achieved, 
\begin{align}
[\hat{A}(r,t), -\epsilon_{0}\hat{E}(r^{'},t)]=\int_{0}^{\infty}\text{d}\omega \dfrac{i\beta(\omega)}{\pi\epsilon_{0}c^{2}\omega\gamma(\omega)}\text{Im}G(r,r^{'},\omega), \label{39}
\end{align}
where $ c $ is the light velocity and $ \gamma(\omega) $ has been introduced in appendix B. By choosing
\begin{equation}
 \beta(\omega)=2\hbar\epsilon_{0}\omega^{2}\gamma(\omega),\label{40}
\end{equation}
 we have, 
\begin{align}
[\hat{A}(r,t), -\epsilon_{0}\hat{E}(r^{'},t)]=\dfrac{2i\hbar}{\pi}\int_{0}^{\infty}\text{d}\omega \dfrac{\omega}{c^{2}}\text{Im}G(r,r^{'},\omega). \label{41}
\end{align}
By considering Eq.\eqref{62a} and Eq.\eqref{63} in appendix A, and using the polar coordinate $ \omega=\vert \omega\vert e^{i\varphi} $, the right hand side of above equation is written as follows, 
\begin{align}
\int_{0}^{\infty}\text{d}\omega \dfrac{\omega}{c^{2}}\text{Im}&G(r,r^{'},\omega)=\dfrac{1}{2i}\int_{-\infty}^{\infty}\text{d}\omega \dfrac{\omega}{c^{2}}G(r,r^{'},\omega)\nonumber\\
=&-\dfrac{1}{2}\text{lim}_{\vert \omega\vert\rightarrow\infty}\int_{0}^{\pi}\text{d}\varphi \dfrac{\omega^{2}}{c^{2}}G(r,r^{'},\omega)\nonumber\\
=&\dfrac{\pi}{2}\delta(r-r^{'}).\label{42}
\end{align}
By substituting \eqref{42} in \eqref{41} the explicit form of canonical commutation relation which is identical with Eq.\eqref{20}, would be obtained, 
\begin{align}
[\hat{A}(r,t), -\epsilon_{0}\hat{E}(r^{'},t)]= i\hbar\delta(r-r^{'}).\label{43}
\end{align}

\subsection{The fluctuation relations}
According to the relation \eqref{16} and the vector potential  operator in \eqref{33},  the electric field  can be written, 
\begin{align}
&\hat{E}^{+}(r,\omega)=i\omega \hat{A}^{+}(r,\omega)=-\omega(\dfrac{ \beta(\omega)}{2k^{"}})^{\frac{1}{2}}\mu_{0}C^{'}\times\nonumber\\
&\lbrace(i\hat{x}+\dfrac{ k_{spp}}{\nu_{i}}\hat{z})e^{(\nu_{i}z)}\Theta(-z)+(i\hat{x}-\dfrac{ k_{spp}}{\nu_{0}}\hat{z})e^{(-\nu_{0}z)}\Theta(z)\rbrace \times\nonumber\\
&[\hat{a}_{R}(x,\omega)+\hat{a}_{L}(x,\omega)]. \label{43a}
\end{align}

 Regarding  to the relation \eqref{21a} and appendix B, the electric field fluctuation can be achieved,
\begin{align}
\langle 0\vert E(r,\omega)E(r^{'},\omega^{'})\vert 0\rangle =2\hbar \omega^{2}\mu_{0}\text{Im}G(r,r^{'},\omega)\delta(\omega-\omega^{'}).\label{44}
\end{align}
On the other hand from \eqref{40} one can deduce the form of the $ \alpha^{m}(\omega) $ and $ \alpha^{d}(\omega) $ in terms of  the media parameters,
\begin{align}
&\vert\alpha^{m}(\omega)\vert =2\hbar \omega^{2}\epsilon_{0}\text{Im}\epsilon_{m}(\omega),\nonumber\\
&\vert\alpha^{d}(\omega)\vert =2\hbar \omega^{2}\epsilon_{0}\text{Im}\epsilon_{d}(\omega).\label{45}
\end{align}
By applying above relations, the noise current fluctuation is given by,
\begin{align}
&\langle 0\vert j(r,\omega)j(r^{'},\omega^{'})\vert 0\rangle =\delta(r-r^{'})\delta(\omega-\omega^{'})\nonumber\\
& 2\hbar \omega^{2}\epsilon_{0}(\text{Im}\epsilon_{m}(\omega)\Theta(-z)+\text{Im}\epsilon_{d}(\omega)\Theta(z)).\label{46}
\end{align}
These results agree with those  calculated by  the fluctuation-dissipation theorem.

\section{The magnetic field variation in amplifying and attenuating media for coherent and squeezed SPP modes}
\subsection{Magnetic field}
In order to express the magnetic field by using \eqref{16} and \eqref{33}, the following relations are necessary, 
\begin{align}
&\dfrac{\partial \hat{a}_{R}(x,\omega)}{\partial x}=ik_{spp}\hat{a}_{R}(x,\omega)+\hat{F}(x,\omega),\nonumber\\
&\dfrac{\partial \hat{a}_{L}(x,\omega)}{\partial x}=-ik_{spp}\hat{a}_{L}(x,\omega)-\hat{F}(x,\omega),\label{43b}
\end{align}
where
\begin{eqnarray}
\hat{F}(x,\omega)&=&(\dfrac{2k^{"}}{\beta(\omega)})^{\frac{1}{2}} \int_{-\infty}^{\infty}\text{d}z^{'}
\Big[(i\hat{x}+\dfrac{ k_{spp}}{\nu_{i}}\hat{z})e^{(\nu_{i}z^{'})}\cdot\hat{j}_{N}^{m+}(x,z,\omega)\Theta(-z)\Theta(-z^{'})\nonumber\\
&+&(i\hat{x}-\dfrac{ k_{spp}}{\nu_{0}}\hat{z})e^{(-\nu_{0}z^{'})}\cdot\hat{j}_{N}^{d+}(x,z,\omega)\Theta(z)\Theta(z^{'})\Big].\label{43c}
\end{eqnarray}
The equations \eqref{43b} are  the well-known quantum Langevin equations. The $ \hat{F}(x,\omega) $ is the operator of Langevin noise source and satisfies the following commutation relation, 
\begin{equation}
[\hat{F}(x,\omega),\hat{F}(x^{'},\omega^{'})]=2k^{"}\delta(x-x^{'})\delta(\omega - \omega^{'}).\label{43d}
\end{equation}
Therefore,  the magnetic field operators of the propagating rightwards SPP modes  are, 
\begin{align}
&\hat{H}^{+}(r,\omega)=-(\dfrac{\beta(\omega)}{2k^{"}})^{\frac{1}{2}}C^{'}\hat{a}_{R}(x,\omega)\times\nonumber\\
&\lbrace (\nu_{i}+\dfrac{ k^{2}_{spp}}{\nu_{i}})e^{(\nu_{i}z)}\Theta(-z)-(\nu_{0}+\dfrac{ k^{2}_{spp}}{\nu_{0}})e^{(-\nu_{0}z)}\Theta(z)\rbrace .\label{43e}
\end{align}
On the other hand, the explicit solution for \eqref{43b} can be derived,
\begin{align}
\hat{a}_{R}(x,\omega)=e^{ik_{spp}(x-x^{'})}\hat{a}_{R}(x^{'},\omega)+\int_{x^{'}}^{x}\hat{F}(y,\omega)e^{ik_{spp}(x-y)}\text{d}y,\label{43f}
\end{align}
where $ x\geq x^{'} $. This equation connects two operators in different  points of space with each other.  By applying the general property of the noise operator $\langle f(x,\omega)\rangle =0$,   the average of  $\hat{a}_{R}(x,\omega)$,   at $x^{'}=0$, the starting point of rightwards propagation, is given by, 
 \begin{equation}
\langle \hat{a}_{R}(x,\omega)\rangle =e^{ik_{spp}x}\langle \hat{a}_{R}(\omega)\rangle .\label{43g}
\end{equation}
The average of the megnetic field  can be then derived,  
\begin{align}
&\langle \hat{H}^{+}(r,\omega)\rangle =-(\dfrac{\beta(\omega)}{2k^{"}})^{\frac{1}{2}}C^{'}e^{ik_{spp}x}\langle \hat{a}_{R}(\omega)\rangle \times\nonumber\\
&\lbrace (\nu_{i}+\dfrac{ k^{2}_{spp}}{\nu_{i}})e^{(\nu_{i}z)}\Theta(-z)-(\nu_{0}+\dfrac{ k^{2}_{spp}}{\nu_{0}})e^{(-\nu_{0}z)}\Theta(z)\rbrace .\label{43h}
\end{align}
\subsection{Propagation in amplifying and attenuating media}
The ohmic loss property of the metal plays the main role in dissipation of  SPP energy and reducing the its propagation length. To overcome these problems one uses a gain dielectric medium (with doped dye molecules) adjust to the metal medium. By this system not only the SPP loss can  be compensated but also the SPP modes can be amplified\cite{32,33,34,35,36}. On the other hand in \eqref{30}, if the imaginary part is positive, then the  SPP modes will be attenuated. In this condition the amplification in the dielectric can not compensate the loss of the metal. In contrast, if it is negative, the SPP modes can be amplified, because the gain of the dielectric can overcome the metal loss \cite{37}. In order to investigate the magnetic field average (see Eq.\eqref{43h}), we consider both attenuating and amplifying systems. Moreover, because of the quantum nature of \eqref{43h} we can consider these conditions for two different kinds of SPP modes, \emph {i.e.}, coherent and squeezed states  \cite{38}.
\subsubsection{coherent state of SPP }
\begin{figure}[ht]
\begin{center}
\includegraphics[scale=0.35]{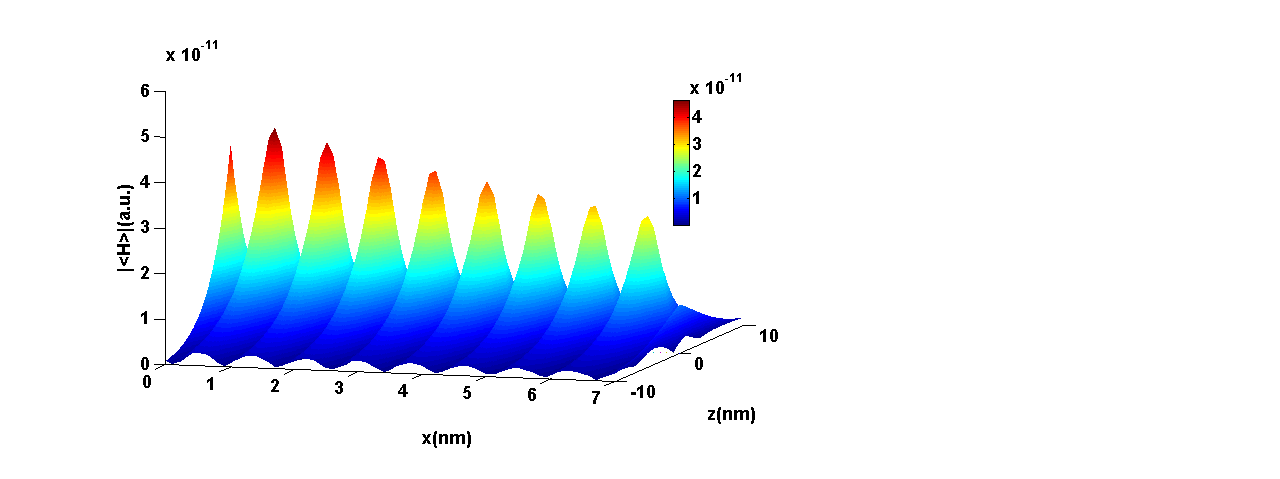}
\end{center}
\caption{The magnetic field propagation of SPP coherent state in  an  attenuating system. The dielectric medium with refractive index $ n_{d}=1.3375-i0.220 $ and silver with $ n_{m}=0.224+i1.34 $ at $ \lambda =350 nm $ are chosen \cite{37}. $ k^{\prime\prime}_{spp}=7.5\times 10^7 $ and $\vert \alpha\vert^{2}=7$.}\label{f2}
\end{figure}

\begin{figure}[ht]
\begin{center}
\includegraphics[scale=0.35]{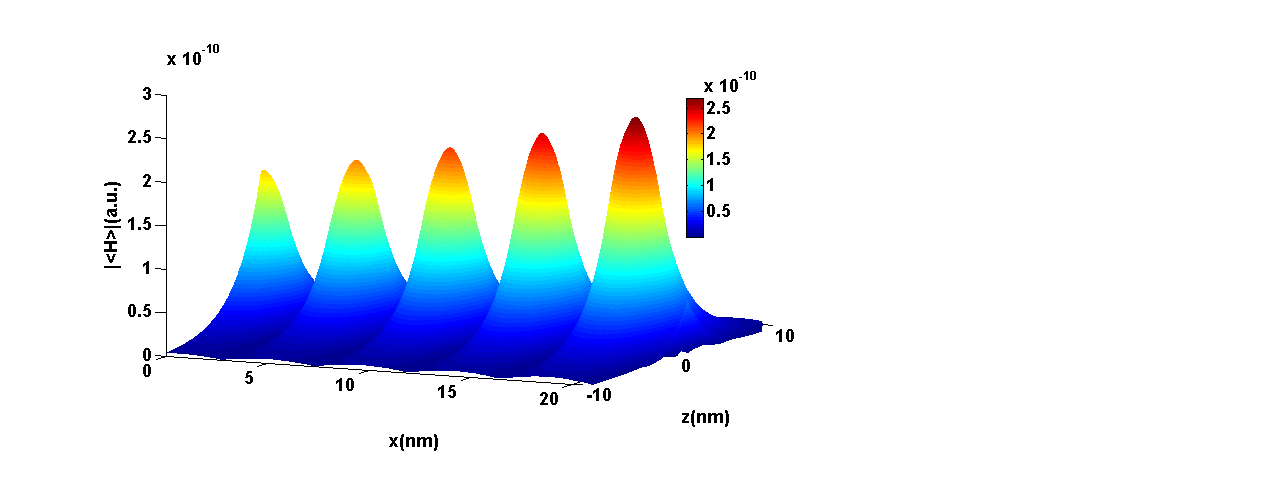}
\end{center}
\caption{The magnetic field propagation of SPP coherent state in an amplifying system. }\label{f3}
\end{figure}

In Ref. \cite{38}, it is mentioned that the SPP operators, like photon operators,  satisfy  the Bosonic commutation relations.  Therefore, we can define the coherent states of SPP as eigenvector of its annihilation operator, in accordance with the definition of the coherent states of radiation field, 
\begin{equation}
\hat{a}_{R}(\omega)\vert\alpha\rangle = \alpha\vert\alpha\rangle .\label{43i}
\end{equation}
By considering the attenuating system \cite{37}  and  the equation \eqref{43i}, the variation of magnetic field average   (see \eqref{43h}) in  coherent state is illustrated in Fig.(\ref{f2}).\\
\\
Fig. (\ref{f2}) shows that for attenuating system, SPP propagation is attenuated because the dielectric gain can not compensate the dissipation in  metal . On the other hand, for dielectric medium with refractive index $n_{d}=1.3375-i0.223 $ the imaginary part of the SPP refractive index is negative and $ k^{\prime\prime}_{spp}=-2.2376\times 10^7 $, so the system can be considered as an amplifying system. Fig.\ref{f3} illustrates SPP magnetic field propagation in the amplifying system. \\
\\
Fig. (\ref{f3}) shows that SPP magnetic field along the propagated axis, can be amplified. It means that the gain of the dielectric has overcompensated the loss in the metal.
\subsubsection{Squeezed states of SPP }
\begin{figure}[ht]
\begin{center}
\includegraphics[scale=0.35]{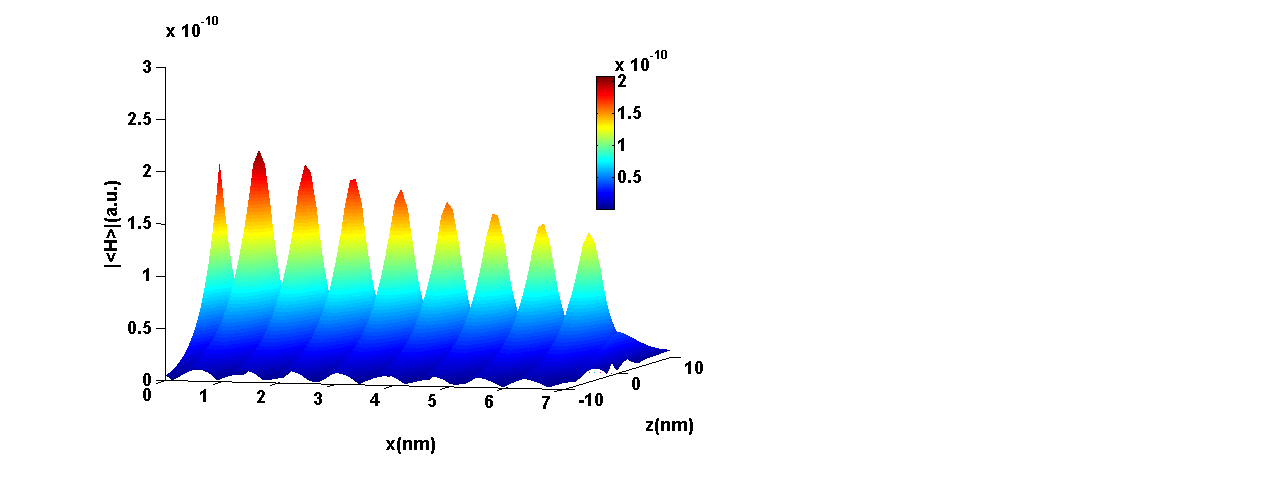}
\end{center}
\caption{ The magnetic field propagation  of SPP squeezed state in an attenuating system. $ \vert \xi\vert =1.5$, $ \theta_{\xi}=0 $, $ \alpha^{'}=\vert \alpha\vert e^{i\theta}  $ and $ \theta =1.5 (rad) $.}\label{f4}
\end{figure}

\begin{figure}[ht]
\begin{center}
\includegraphics[scale=0.35]{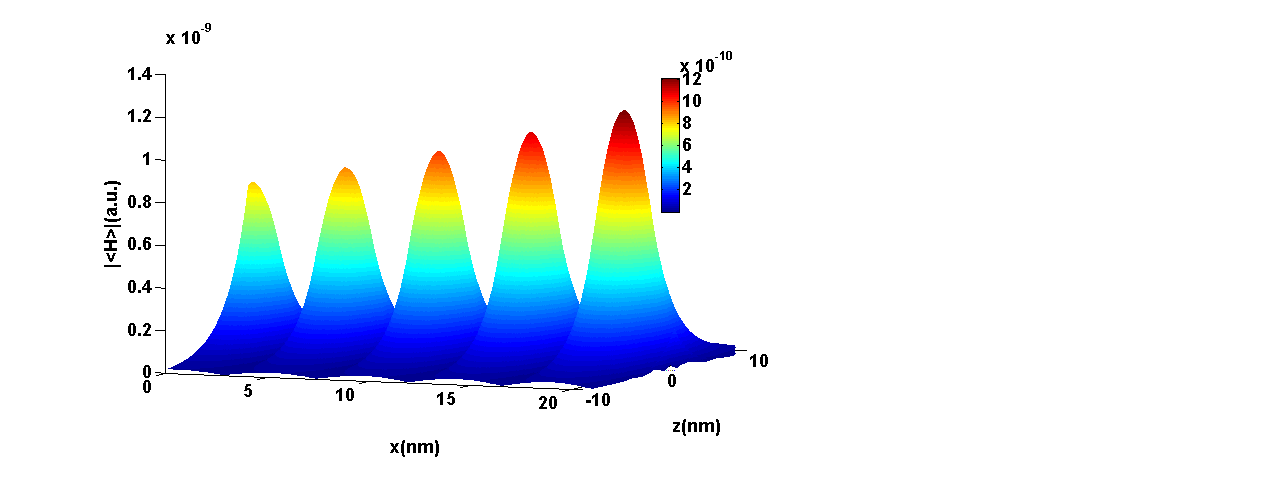}
\end{center}
\caption{The magnetic field propagation of  SPP squeezed state in an  amplifying system.}\label{f5}
\end{figure}

After observing the quadrature squeezing of SPP in a gold waveguide \cite{39}, the quantum mechanical description of these states have been given \cite{38}. The squeezed state of SPP  is generated, similar to the  squeezed state  of radiation field, by applying the  squeezed operator on vacuum, 
\begin{equation}
|\xi\rangle=\hat{s}(\xi)|0\rangle=\exp(\dfrac{1}{2}\xi^{*}\hat{a}^{2}-\dfrac{1}{2}\xi\hat{a}^{\dagger^{2}})|0\rangle,\label{43g}
\end{equation}
where $ \xi=\vert \xi\vert e^{i\theta_{\xi}} $, $ \vert \xi\vert $ and $ \theta_{\xi} $ are the squeezed parameter and compressed angle, respectively. If the SPP is prepared in the coherent  squeezed state $ \vert \xi , \alpha^{'}\rangle $, then
\begin{equation}
\hat{a}(\omega)\vert \xi , \alpha^{'}\rangle =(\mu {\alpha^{\prime}}(\omega)-\nu {\alpha^{\prime}}(\omega))\vert\alpha^{'}\rangle,\label{43k}
\end{equation} 
where $ \mu=\cosh( \vert \xi\vert)$ and $ \nu= \sinh( \vert \xi\vert) e^{i\theta_{\xi}}$.
By applying  \eqref{43h} and \eqref{43k}, we can illustrate the magnetic field average of SPP  squeezed state, as given  for an attenuating system in Fig. (\ref{f4}).\\
\\ 
This figure shows that, the propagation of the SPP will undergo more  attenuation in any state other than squeezed state, e.g., if the initial SPP state is prepared in squeezed state, the amplitude of the magnetic field is greater than the  coherent state case (see Fig.\ref{f2}).\\
Likewise, the SPP magnetic field in an amplifying system and squeezed state  is shown in Fig.\ref{f5}.\\
\\ 
The comparison between Fig. (\ref{f5}) and Fig. (\ref{f3}) shows that  those  systems that  the SPP modes is amplified, the magnitude of magnetic field in squeezed state is several times greater than  in the coherent state case.
\subsubsection{comparison between SPP coherent and squeezed state in the amplifying system}
\begin{figure}[ht]
\begin{center}
\includegraphics[scale=0.30]{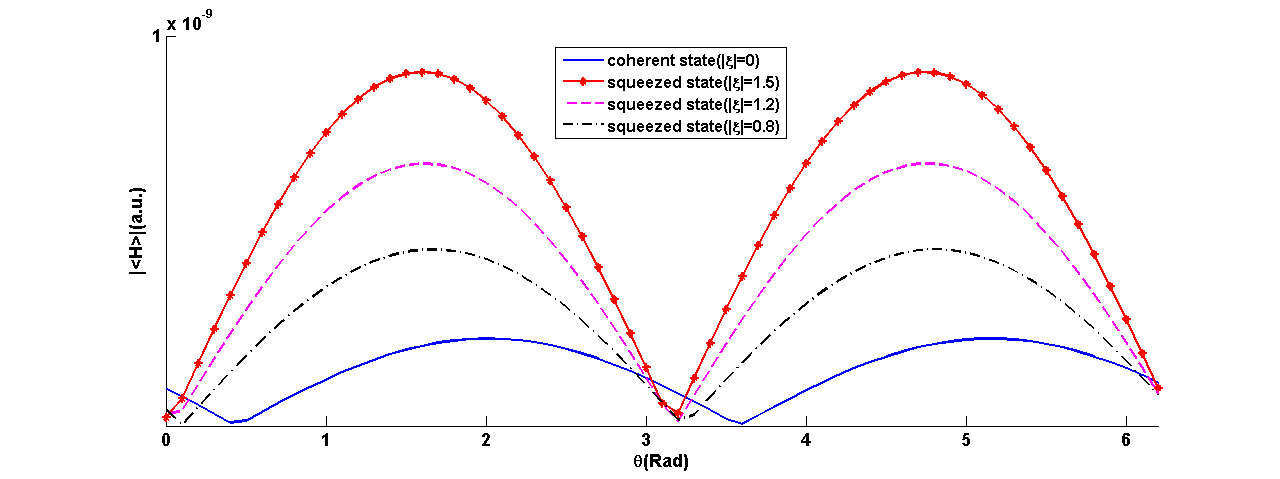}
\end{center}
\caption{The variation of SPP magnetic field average  for different phases and different values of squeezed parameter,  $ z=0 $ and $ x=10 nm $. }\label{f6}
\end{figure}

\begin{figure}[ht]
\begin{center}
\includegraphics[scale=0.35]{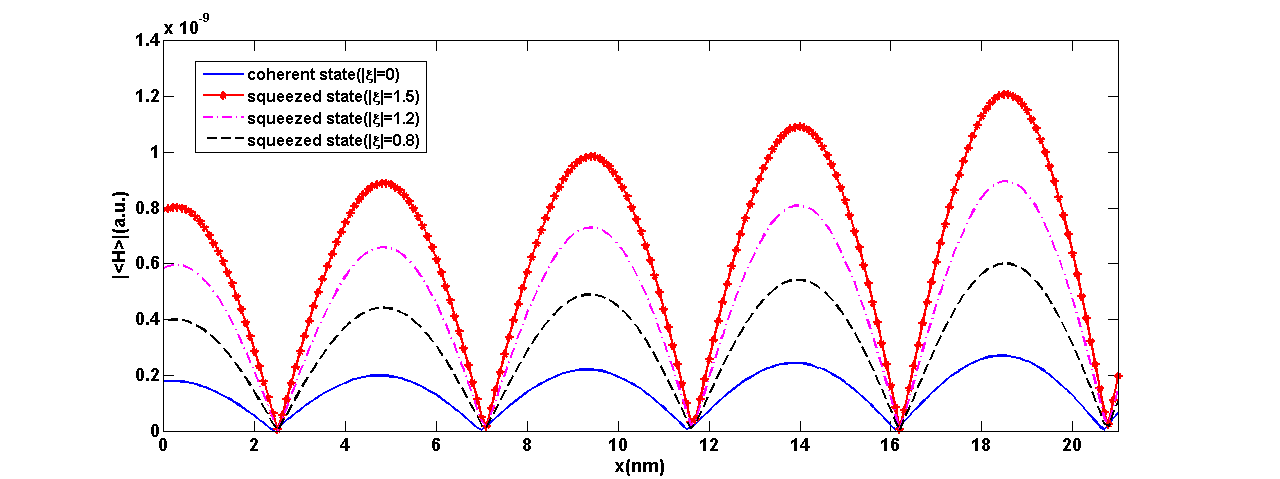}
\end{center}
\caption{The variation of  SPP magnetic field average  for different values  of squeezed parameter,  $ z=0 $ and $ \theta=1.5 Rad $. }\label{f7}
\end{figure}

In  Ref. \cite{40} the importance of the phase on the properties of a  squeezed state, has been studied. Accordingly we can investigate the influence of the phase by considering the variation of the average of  SPP magnetic field in \eqref{43h} and \eqref{43k} versus  $ \theta $, where $\theta$ is the argument of $\alpha^{\prime}=|\alpha^{\prime}|e^{i\theta}$.  It is depicted  in Fig. (\ref{f6}). \\
\\ 
As is illustrated in Fig. (\ref{f6}),  the SPP magnetic field in squeezed state is greater than in the case of coherent state. The drastic  difference between coherent and squeezed state is occurred  in $ \theta= 1.5 Rad $. In Fig. (\ref{f7}) we demonstrate the SPP magnetic field for various parameters of squeezed states.\\
\\ 
Fig. (\ref{f7}) shows that by choosing an appropriate phase, the magnetic of  SPP in squeezed state is several times greater than that of coherent state.
\section{conclusion}
The quantization of SPP by a new method, based on the Green's tensor, provides the possibility to distinguish the behavior of different modes of the quantized field, such as coherent and squeezed states of SPP. The propagation of  different modes of  quantized  SPP can also be investigated in  amplifying and attenuating media. We have seen that the  behavior of these two modes  in the said media are  distinguished  drastically. \\
We think,  by extension of  this method, the quantization of SPP in different geometries, other than the plane interfaces,  can  also be done straightforwardly, and accordingly it leads to  understand  the more interesting properties of SPP. 

\appendix
\section*{appendix A}
The Green's tensor satisfy the maxwell's equation when the source term is replaced  by Dirac delta function( see Eq. \eqref{19}):
\begin{equation}
-\nabla\times\nabla\times G(r,r^{'},\omega)+\dfrac{\omega^{2}}{c^{2}}\epsilon(r,\omega)G(r,r^{'},\omega)=I \delta(r-r^{'}).\label{62}
\end{equation} 
For large frequency, $ \epsilon(r,\omega)\rightarrow 1 $ and from above equation one can deduce that:
\begin{equation}
\lim_{\vert \omega\vert\rightarrow\infty}\dfrac{\omega^{2}}{c^{2}}G(r,r^{'},\omega)=-\delta(r-r^{'})\label{62a}
\end{equation}
The Green's tensor has also general properties such as:
\begin{align}
& G^{T}(r,r^{'},\omega)=G(r^{'},r,\omega)\nonumber\\
&G^{*}(r,r^{'},\omega)=G(r,r^{'},-\omega)\label{63}
\end{align}
On the other hand for a nonmagnetic media,by some algebra, one can obtain a very usefull relation\cite{31}
\begin{equation}
\int \text{d}s \text{Im}\epsilon(s,\omega)G(r,s,\omega) \cdot G^{*}(s,r^{'},\omega)=\dfrac{c^{2}}{\omega^{2}}\text{Im}G(r,r^{'},\omega)\label{64}
\end{equation} 
\section*{appendix B}
The right hand side of Eq.\eqref{38} can be express versus Green's tensor. By considering Eq.\eqref{21b} and the Green's tensor for a metal-dielectric geometry Eq. \eqref{31}, one can prove that:
\begin{align}
&\int \text{d}s \text{Im}\epsilon(s,\omega)G(r,s,\omega) \cdot G^{*}(s,r^{'},\omega)=\nonumber\\
&\dfrac{\vert k_{spp}\vert^{2}}{2k^{'}k^{"}}\gamma(\omega) \vert C^{'}\vert^{2}\lbrace\dfrac{e^{ik_{spp}\vert x-x^{'}\vert}}{k_{spp}}+\dfrac{e^{-ik^{*}_{spp}\vert x-x^{'}\vert}}{k^{*}_{spp}}\rbrace\times\nonumber\\
&[(i\hat{x}+\dfrac{ k_{spp}}{\nu_{i}}\hat{z})e^{(\nu_{i}z)}\Theta(-z)+(i\hat{x}-\dfrac{ k_{spp}}{\nu_{0}}\hat{z})e^{(-\nu_{0}z)}\Theta(z)] \times\nonumber\\
&[(-i\hat{x}+\dfrac{ k^{*}_{spp}}{\nu^{*}_{i}}\hat{z})e^{\nu^{*}_{i}z^{'}}\Theta(-z^{'})+(-i\hat{x}-\dfrac{ k^{*}_{spp}}{\nu^{*}_{0}}\hat{z})e^{-\nu^{*}_{0}z^{'}}\Theta(z^{'})],\label{65}
\end{align} 
where
\begin{equation}
\gamma(\omega)=(1+\dfrac{\vert k_{spp}\vert^{2}}{\vert\nu_{i}\vert^{2}})\dfrac{\text{Im}\epsilon_{m}}{\nu_{i}+\nu_{i}^{*}}+(1+\dfrac{\vert k_{spp}\vert^{2}}{\vert\nu_{0}\vert^{2}})\dfrac{\text{Im}\epsilon_{d}}{\nu_{0}+\nu_{0}^{*}}.\label{66}
\end{equation}
Equating Eq.\eqref{38} and Eq.\eqref{65} and considering Eq.\eqref{64} leads to:
\begin{align*}
&[\hat{A}(r,t), -\epsilon_{0}\hat{E}(r^{'},t)]=\int \text{d}\omega \dfrac{i\omega\epsilon_{0}\mu_{0}^{2}\beta(\omega)}{\pi \gamma(\omega)}\dfrac{c^{2}}{\omega^{2}}\text{Im}G(r,r^{'},\omega).
\end{align*} 

\end{document}